%% file: draftCPF.tex
\documentclass[aps,prd,twocolumn,showpacs,superscriptaddress,groupedaddress]{revtex4-2}  
\usepackage{graphicx}  
\usepackage{dcolumn}   
\usepackage{bm}        
\usepackage{amssymb}   

\usepackage{lineno}
\usepackage{booktabs}
\usepackage{longtable}
\usepackage{overpic}
\usepackage{multirow}
\usepackage{amsmath}
\usepackage{color}
\usepackage[bookmarksnumbered, pdfstartview=FitH,colorlinks,urlcolor=blue, citecolor=blue,linkcolor=blue]{hyperref}

\usepackage{subfigure}
\usepackage{tabularx}
\usepackage{textcomp}
\usepackage{gensymb}
\usepackage[capitalise]{cleveref}

\let\oldequation\equation
\let\oldendequation\endequation

\renewenvironment{equation}
  {\linenomathNonumbers\oldequation}
  {\oldendequation\endlinenomath}

\crefname{section}{Sec.}{Secs.}
\begin{document}



\title{Determination of the \texorpdfstring{$C\!P$}{CP}-even fraction
of \texorpdfstring{$D^0\rightarrow K_S^0\pi^+\pi^-\pi^0$}{Ks3pi}}
\author{
\begin{small}
\begin{center}
  \input{./authorlist_2023-01-18}
\end{center}
\vspace{0.4cm}
\end{small}
}


\begin{abstract}
  Quantum-correlated $D\bar{D}$ pairs collected by the
BESIII experiment at the $\psi(3770)$ resonance, corresponding to an
integrated luminosity of 2.93 fb$^{-1}$, are used to study the $D^0 \rightarrow
K^{0}_S\pi^{+} \pi^{-} \pi^{0}$ decay mode. The $C\!P$-even fraction of  $D^0 \rightarrow
K^{0}_S\pi^{+} \pi^{-} \pi^{0}$ decays is
determined to be $0.235\pm 0.010\pm 0.002$, where the first uncertainty is statistical
and the second is systematic.
\end{abstract}


\maketitle

\section{Introduction}\label{sec:intro}
The complex phase of the Cabibbo-Kobayashi-Maskawa~(CKM) matrix is
the only source of the $C\!P$ violation within the quark sector of the
Standard Model~(SM)~\cite{Hocker:2006xb}. However, this source of $C\!P$ violation is too
weak to explain the huge matter-antimatter asymmetry in the
universe. This fact motivates precise studies of flavor transitions to test the CKM paradigm and to search
for evidence of $C\!P$ violation beyond the SM.  The unitarity triangle is a graphical representation of
the CKM matrix in the complex plane, with angles denoted  $\alpha$, $\beta$, and
$\gamma$~\cite{Workman:2022ynf}. An improved measurement of the angle $\gamma$ is of particular 
importance in studies of $C\!P$ violation. Comparisons between the values of $\gamma$ obtained from direct
measurements~\cite{Workman:2022ynf} and global CKM
fits~\cite{Charles:2015gya} provide an important test of CKM unitarity and
allow for searches for new physics beyond the SM.

The angle $\gamma$ of the unitarity triangle is measured in decays which are sensitive to
interference between favored $b\to c$ and suppressed $b\to u$ quark
transition amplitudes~\cite{LHCb:2021dcr}. Typically, the interference is
measured in $B^\pm \rightarrow D h^{\pm}$ decays, where $D$ is an
admixture of $D^0$ and $\bar{D}^0$ flavor states and $h^\pm$ is either a
charged pion or kaon. The theoretical
uncertainty of the measurement of $\gamma$ through this approach is negligible~\cite{Brod:2013sga}. The current uncertainty of
the $\gamma$ measurement is statistically
dominated~\cite{LHCb:2021dcr}. Therefore, including more $D$-decay modes
is desirable to improve the precision of the $\gamma$ measurement. One of the
important classes of  $B^\pm \rightarrow D h^{\pm}$ measurement is the so-called `GLW' strategy, in which the $D$ decay is reconstructed in a $C\!P$ eigenstate~\cite{Gronau:1991dp}. 
This strategy can be extended to encompass quasi-$C\!P$ eigenstates that are predominantly $C\!P$ odd or even~\cite{Malde:2015mha}. This approach requires that the $C\!P$-even fraction, quantified by the parameter $F_+$, in known.  A promising quasi-$C\!P$ eigenstate is the decay 
$D\rightarrow K_S^0\pi^+\pi^-\pi^0$, which has a higher branching fraction than decays to individual $C\!P$ eigenstates and is known to be predominantly $C\!P$ odd from a measurement performed with data collected by CLEO-c corresponding to an integrated luminosity of 0.82~fb$^{-1}$ that yielded the result  $F_+ = 0.238 \pm 0.012\pm 0.012$~\cite{Resmi:2017fuo}.  In this paper we present an improved determination of this parameter made with quantum-correlated $D\bar{D}$ pairs, based
on a data sample of 2.93 fb$^{-1}$ taken at the $\psi(3770)$ resonance by the
BESIII detector.

\section{Formalism}%
\label{sec:Formulism}

The wave function for the pairs of neutral charm mesons produced at the
$\psi(3770)$ resonance is anti-symmetric because of their odd charge conjugation. This quantum correlation allows the
$C\!P$-even fraction $F_+$ of $D^0\rightarrow K_S^0\pi^+\pi^-\pi^0$ decays to
be determined with double-tag~(DT)~\cite{Li:2021iwf} yields, in which the $D$
meson is reconstructed in the signal decay and the $\bar{D}$ meson is
reconstructed in one of several tag modes~\footnote{Throughout this paper the
$C\!P$ conjugated process is implicit.}.
 The tag modes used in this analysis are
summarized in Table~\ref{tab:tagmodes}.
\begin{table}[b]
  \centering
  \caption{The tag modes used in this analysis.}
  \label{tab:tagmodes}
  \begin{ruledtabular}
    \begin{tabular}{cc}
      Type & Tag modes \\
      \hline
      $C\!P$-even & $K^{+} K^{-}, \pi^{+} \pi^{-}, K_{{S}}^{0} \pi^{0} \pi^{0}, K_{{L}}^{0} \omega, K_{{L}}^{0} \pi^{0}$ \\
      $C\!P$-odd & $K_{{S}}^{0} \pi^{0}, K_{{S}}^{0}
      \eta(\gamma\gamma), K_{{S}}^{0} \eta^{\prime}(\eta\pi^+\pi^-),
      K_{{S}}^{0} \eta^{\prime}(\gamma\pi^+\pi^-)$  \\
      Quasi-$C\!P$ & $\pi^{+} \pi^{-} \pi^{0},\pi^{+} \pi^{-} \pi^+\pi^-$ \\
      Mixed $C\!P$ & $ K_{{S,L}}^{0} \pi^{+} \pi^{-}$ \\
      Self-tag & $K_{{S}}^{0} \pi^{+} \pi^{-} \pi^0$ \\
    \end{tabular}
  \end{ruledtabular}
\end{table}

The expected DT yield is given
by~\cite{Malde:2015mha}
\begin{equation}
  \label{eq:dtFp}
  \begin{aligned}
    N_\mathrm{DT}  =&  2N_{D \bar{D}}\mathcal{B} (S) \mathcal{B} (T)\varepsilon(S|T)\\
  &   \times [1 - (2 F_+ - 1)  (2 F_+^T - 1)],
  \end{aligned}
\end{equation}
where $N_{D \bar{D}}$ is the number of $D^0\bar{D}^0$
pairs~\cite{BESIII:2018iev} produced at the $\psi(3770)$ resonance,
$S$~($T$ ) denotes the signal mode $D\to K^0_S\pi^+\pi^-\pi^0$~(tag mode),
$\mathcal{B}(S)$~[$\mathcal{B}(T)$] is the branching fraction of the $D^0$
decaying into the $S$~($T$), $\varepsilon(S|T)$ is
the DT efficiency, $F_+$~($F_+^T$)
is the $C\!P$-even fractions for the $D\rightarrow S$~($D\rightarrow T$)
decays. In addition, \cref{eq:dtFp} omits terms of order $\mathcal{O}(x_D^2,
y_D^2)\sim10^{-5}$ associated with $D^0\bar{D}^0$ mixing effects, where $x_D$
and $y_D$ are the mixing parameters~\cite{Workman:2022ynf}.
The dependency on the $N_{D \bar{D}}$,
$\mathcal{B} (T)$, and single-tag~(ST) efficiency of the tag mode is removed by measuring the
ST yield of the tag mode $T$.
The expected ST yield for the tag mode $T$ is given by~\cite{Malde:2015mha}
\begin{equation}
  \label{eq:styields}
  N_\text{ST}(T)=2 N_{D \bar{D}} \mathcal{B}(T) \varepsilon(T),
\end{equation}
where $\varepsilon(T)$ is the ST efficiency.  A further correction of
$1 /\left[1-(2F_+^T-1) y_D\right]$ is applied to  $N_\mathrm{ST}$ to account for
$D^0\bar{D}^0$ mixing effects, where $y_D$ is one of the
$D^0\bar{D}^0$ mixing parameters~\cite{Workman:2022ynf}. With assumption of
\mbox{$\varepsilon(S|T) = \varepsilon(S) \cdot\varepsilon(T)$}, the ratio of
$N_\mathrm{DT}$ to $N_\mathrm{ST}$ can be written as
\begin{equation}
  \label{eq:norDT}
  R=\mathcal{B}(S) \varepsilon(S)\left[1-\left(2
F_+-1\right)\left(2 F_+^T - 1\right)\right].
\end{equation}
For $C\!P$-even~($C\!P$-odd) tags, $F_+^{T}$ is equal to 1~(0). Then the
corresponding ratio $R^-$~($R^+$) for the $C\!P$-tag
mode $T$ is given by
\begin{equation}
  \label{eq:norDT_CP}
  R^\mp=\mathcal{B}(S) \varepsilon(S)\left[1-\eta_{T}^{\pm}\left(2
  F_+-1\right)\right],
\end{equation}
where $\eta_T^+$~($\eta_T^-$) is the $C\!P$ eigenvalue of the $C\!P$-even~($C\!P$-odd) tag
mode $T$. With such $C\!P$-eigenstate tags, the $C\!P$-even fraction is
given by 
\begin{equation} \label{eq:fpCP}
  F_+=\frac{R^+}{R^++R^-},
\end{equation}
which has no dependence on $\mathcal{B}(S)$ and $\varepsilon(S)$. For
a quasi-$C\!P$-tag mode $T$ of known $C\!P$-even fraction $F_+^T$, the $C\!P$-even fraction of the signal mode is
\begin{equation}
  \label{eq:FpMixedCP} 
  F_+ = \frac{R^+ F_+^T}{R^T - R^+ + 2 R^+ F_+^T},
\end{equation}
where $R^T$ is the ratio of the measured DT yield to the measured ST yield for the tag mode $T$.
The signal decay $D\to K_S^0\pi^+\pi^-\pi^0$ can also be used as
a self-tag mode. Denoting the ratio of the DT yield of the self-tag mode to the
corresponding ST yield by $R^S$, the $C\!P$-even fraction is given by 
\begin{equation}
  \label{eq:fpks3pi}
  F_+=\frac{R^{S}}{R^-}.
\end{equation}

\begin{figure}[b]
  \centering
  \includegraphics[width=0.47\textwidth]{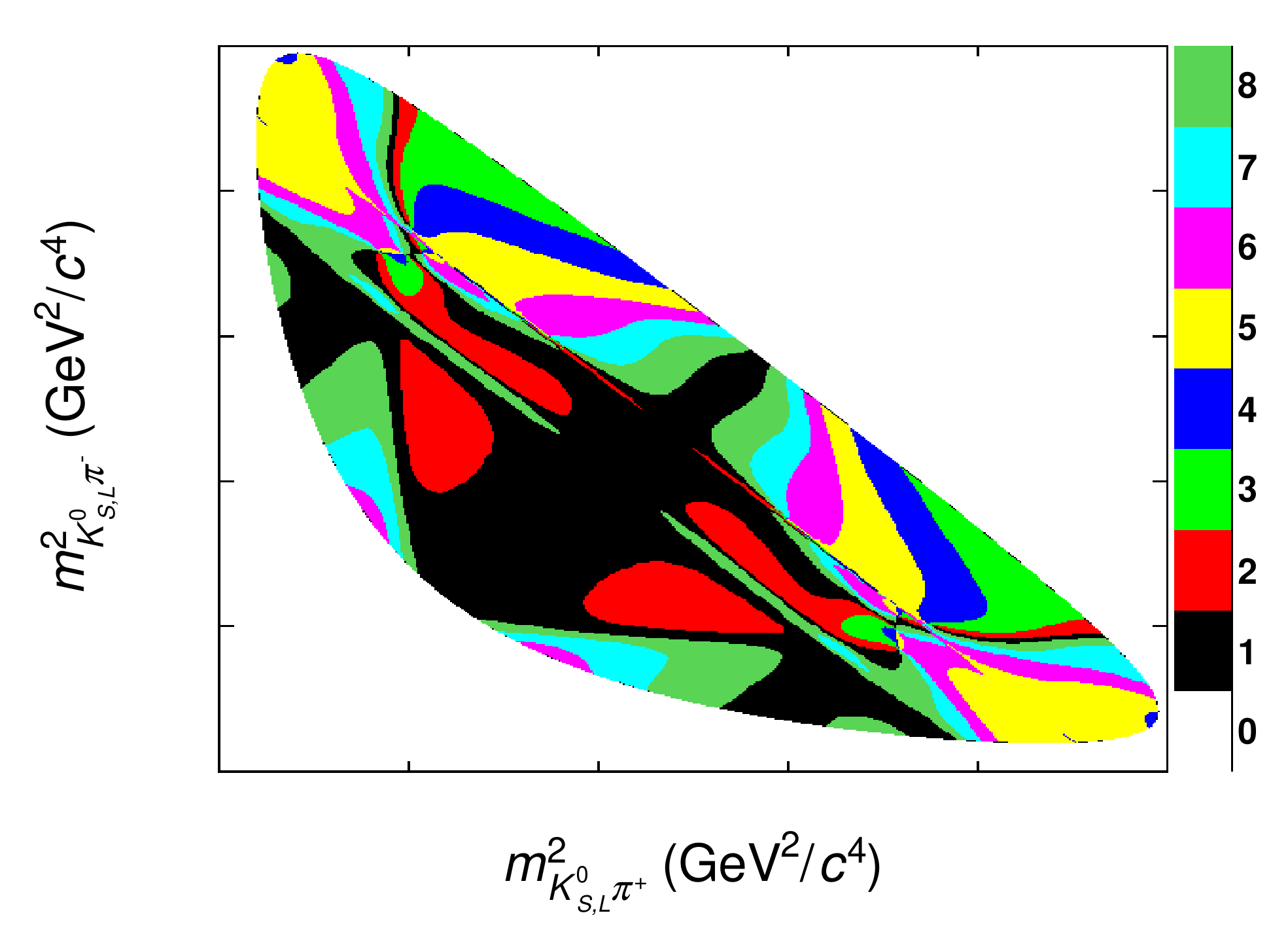}
  \caption{The binning of the $D\rightarrow
  K_{S,L}^0\pi^+\pi^-$ Dalitz plot. The
color scale represents the absolute value of the bin number $|i|$, where $i$
is negative for the bin with $m_{K_{S,L}^0\pi^+}<m_{K_{S,L}^0\pi^-}$.}
\label{fig:binningKspipi}
\end{figure}

There is no direct measurement of $F_+$ for the mixed $C\!P$-tag modes
$D\to K_{S,L}^0\pi^+\pi^-$.  However, quantum-correlated studies indicate that $F_+^{K_{S,L}^0\pi^+\pi^-}$ is close to 0.5 for both decay modes~\cite{BESIII:2020khq}. It follows from  \cref{eq:dtFp} that using these decays integrated over all phase space as tags gives very low sensitivity to $F_+$ of the signal mode.
 Therefore a localized measurement is pursued, in which the 
DT yields are measured in the pairs of eight bins in the Dalitz plot of
$D\rightarrow K_{S,L}\pi^+\pi^-$.  The binning scheme adopted is the 
``equal $\Delta\delta$ binning" of Ref.~\cite{CLEO:2010iul},
as shown as \cref{fig:binningKspipi}. The
expected DT yield with the $K_{S}^0\pi^+\pi^-$ final state
in the $i$-th bin can be written as~\cite{BESIII:2020khq}
\begin{equation}
  \label{eq:fpKspipi}
  \begin{aligned}
    M_i  = h \left[K_i + K_{- i}  - 2 c_i  \sqrt{K_i K_{- i}}  (2 F_+ - 1) \right]
  \end{aligned}
\end{equation}
and the corresponding expected yield for the $K_{L}^0\pi^+\pi^-$ case as 
\begin{equation}
  \label{eq:fpKlpipi}
  \begin{aligned}
    M'_i  = h'  \left[K_i' + K_{- i}'  + 2 c_i'  \sqrt{K_i' K_{- i}'}  (2 F_+ - 1) \right],
  \end{aligned}
\end{equation}
where for the decay $D \to K_{S}^0\pi^+\pi^-$ ($D \to K_{L}^0\pi^+\pi^-$) $h$~($h'$) is a normalization factor, $K_i$~($K_i'$) is the flavor-tagged fraction of $D^0\rightarrow
K_{S}^0\pi^+\pi^-$~($D^0\rightarrow K_L^0\pi^+\pi^-$) with
the final state in the $i$-th bin, and $c_i$~($c'_i$) is the
amplitude-weighted average of cosine of the strong-phase
difference~\cite{BESIII:2020khq}. An event can migrate between the bins
because of the finite detector resolution. To improve the 
momentum resolution of the final-state particles and suppress the migration between phase-space bins, the invariant mass of the
$K_S^0\pi^+\pi^-$ final state is constrained to the known $D^0$
mass~\cite{Workman:2022ynf}, denoted as $M_\mathrm{PDG}^{D}$.

\section{Detector and data samples}
\label{sec:detSample}
The BESIII detector~\cite{BESIII:2009fln} records symmetric $e^+e^-$
collisions provided by the BEPCII storage ring~\cite{Yu:2016cof} in the
center-of-mass energy range from 2.0 to 4.95~GeV, with a peak luminosity of $1
\times 10^{33}\;\text{cm}^{-2}\text{s}^{-1}$ achieved at center-of-mass energy $\sqrt{s} =
3.77\;\text{GeV}$. BESIII has collected large data samples in this energy
region~\cite{BESIII:2020nme}. The cylindrical core of the BESIII detector
covers 93\% of the full solid angle and consists of a helium-based multilayer
drift chamber~(MDC), a plastic scintillator time-of-flight system~(TOF), and a
CsI(Tl) electromagnetic calorimeter~(EMC), which are all enclosed in a
superconducting solenoidal magnet providing a 1.0~T magnetic field~\cite{Huang:2022wuo}. The 
solenoid is supported by an octagonal flux-return yoke with resistive plate
counter muon identification modules interleaved with steel. The
charged-particle momentum resolution at $1~{\rm GeV}/c$ is $0.5\%$, and the
${\rm d}E/{\rm d}x$ resolution is $6\%$ for electrons from Bhabha scattering.
The EMC measures photon energies with a resolution of $2.5\%$ ($5\%$) at
$1$~GeV in the barrel (end-cap) region. The time resolution in the TOF barrel
region is 68~ps, while that in the end-cap region is 110~ps.

Simulated data samples produced with a {\sc
geant4}-based~\cite{GEANT4:2002zbu} MC package, which includes
the geometric description of the BESIII detector and the detector response,
are used to determine detection efficiencies and to estimate backgrounds. The
simulation models the beam-energy spread and initial-state radiation (ISR) in
the $e^+e^-$ annihilations with the generator {\sc kkmc}~\cite{Jadach:2000ir}.
The inclusive MC sample includes the production of $D\bar{D}$ pairs, the
non-$D\bar{D}$ decays of the $\psi(3770)$, the ISR production of the $J/\psi$
and $\psi(3686)$ states, and the continuum processes incorporated in {\sc
kkmc}~\cite{Jadach:2000ir}. All particle decays are modeled with {\sc
evtgen}~\cite{Lange:2001uf, *Ping:2008zz_BesGen} using branching fractions
taken from the PDG~\cite{Workman:2022ynf}. Final-state
radiation from charged final state particles is incorporated using the
{\sc photos} package~\cite{Richter-Was:1992hxq}. For each tag mode, signal MC
samples for ST and DT are produced. The multibody decay is
generated with available amplitude modes or the expected intermediate
resonance involved.  

\section{Event Selection}
\label{:sec:evetSelect}

Charged tracks detected in the MDC are required to be within a polar
angle ($\theta$) range of $|\!\cos\theta|<0.93$, where $\theta$ is
defined with respect to the $z$-axis, which is the symmetry axis of
the MDC. For charged tracks not originating from $K_S^0$ decays, the
distance of the closest approach to the interaction point must be
less than 10\,cm along the $z$-axis, $|V_{z}|$,  and less than 1\,cm
in the transverse plane, $|V_{xy}|$. Particle identification~(PID)
for charged tracks combines measurements of the specific ionization energy loss in
the MDC~(d$E$/d$x$) and the flight time in the TOF to form
likelihoods $\mathcal{L}(h)~(h=K \text{~and~}\pi)$ for each hadron $h$
hypothesis. Charged kaons and pions are
identified by comparing the likelihoods for kaon and pion
hypotheses, $\mathcal{L}(K)>\mathcal{L}(\pi)$ and
$\mathcal{L}(\pi)>\mathcal{L}(K)$, respectively.

Photon candidates are identified using showers in the EMC.  The
deposited energy of each shower must be more than 25~MeV in the
barrel region ($|\!\cos \theta|<0.80$) and more than 50~MeV in the end-cap region ($0.86 <|\!\cos \theta|< 0.92$). To suppress
electronic noise and showers unrelated to the event, the difference
between the EMC time and the event start time is required to be
within [0, 700]\,ns.

Each $K_{S}^0$ candidate is reconstructed from two oppositely charged
tracks satisfying $|V_{z}|<$ 20~cm. The two charged tracks are assigned
as $\pi^+\pi^-$ without imposing further PID criteria. They are constrained to
originate from a common vertex and are required to have an invariant mass
within [0.485, 0.510]~GeV/$c^2$. The decay length of the $K^0_S$
candidate is required to be greater than twice the vertex resolution
away from the interaction point.

Pairs of selected photon candidates are used to reconstruct
$\pi^0$~($\eta$) candidates. The invariant masses of the photon pairs
are required to be within $[0.110, 0.155]$~([0.480, 0.580]) GeV/$c^2$. To
improve the momentum resolution, a kinematic fit is applied to
constrain the $\gamma\gamma$ invariant mass to the known 
$\pi^0$($\eta$) mass~\cite{Workman:2022ynf}, and the
$\chi^2$ of the kinematic fit is required to be less than 20. The
fitted momenta of the $\pi^0$($\eta$) are used in the subsequent 
analysis. The $\omega$ candidates are selected by requiring the
invariant mass of the $\pi^+\pi^-\pi^0$ combination to be within [0.750, 0.820]~GeV/$c^2$. The $\eta^{\prime}\rightarrow \pi^+\pi^-\eta
$ and $\eta ^{\prime}\rightarrow \gamma\pi^+\pi^-$ decays are used to
reconstruct $\eta ^\prime$ mesons, with the invariant masses of the
$\pi^+\pi^-\eta $ and $\pi^+\pi^-\gamma$ required to lie within
[0.938, 0.978]~GeV/$c^2$.

For the $D\rightarrow K^{+} K^{-}$ and $D\rightarrow\pi^{+} \pi^{-}$
tag modes, the background from cosmic rays and Bhabha events are
suppressed with the following requirements~\cite{BESIII:2014rtm}.
First, the two charged tracks used as the $C\!P$ tag must have a TOF time
difference less than 5 ns and they must not be consistent with being a muon
pair or an electron-positron pair. To be a muon candidate, the track must have
$|\chi_{\text{d}E/\text{d}x}| < 5$,
$0.15 < E_\text{EMC} < 0.30$ GeV, and $d_\mu > 40$ or $d_\mu > 80 p - 60$,
where $\chi_{\text{d}E/\text{d}x}$~\cite{Cao:2010zzd} is the $\chi$ for the
hypothesis of muon in the PID with d$E$/d$x$, $E_\text{EMC}$ is deposited
energy in the EMC by the track, $d_\mu$ is the penetration depth in the muon
identification modules with unit of center meter, and $p$ is the momentum of the track. To be an
electron~(positron) candidate, the track must have greater probability
(combing the d$E$/d$x$ and flight time) being an electron~(positron) than
being a kaon or pion. If the corresponding EMC shower is reconstructed, the
track must satisfy: (1) $E/p > 0.8$ if $|\!\cos \theta|<0.70$ , or (2) $E/p >
-7.5 |\!\cos \theta| + 6.05$ if $0.7 < |\!\cos \theta| < 0.8$, or (3) $E/p >
0.6$ if $|\!\cos \theta| > 0.85$, where $E/p$ is the ratio of the deposited
energy in the EMC to the momentum of the track and $\theta$ is the polar angle
of the shower. Second, there must be at least one EMC
shower (other than those from the $C\!P$-tag tracks) with an energy larger than 50~MeV or at least one additional charged track detected in the MDC.
For the $D\rightarrow \pi^+\pi^-\pi^0$ and $D\rightarrow \pi^+\pi^- \pi^+
\pi^-$ tag modes,  events
are rejected if any combination of $m_{\pi^+\pi^-}$ lies in the mass
window of $[0.481, 0.514]$~GeV/$c^2$ in order to suppress the background from $K_S^0$ decays. 

\section{ Determination of the ST yield}\label{sec:styields}

The tag modes that do not involve a $K_L^0$ are fully reconstructed. The signal
candidate of the fully reconstructed tag mode is identified by the
beam-energy constrained mass
\begin{equation}
  \label{eq:mbc}
  M_\mathrm{BC}=\sqrt{E_\mathrm{beam}^2-p_D^2},
\end{equation}
where $E_\text{beam}$ and $p_D$ are the beam energy and reconstructed momentum
of the $D$ candidate in the $e^+e^-$ center-of-mass frame, respectively. If multiple
combinations are reconstructed for an event, the combination with the lowest value of 
$|\Delta E|$ is retained for further analysis, where $\Delta E$ is the difference
between the $E_\mathrm{beam}$ and the reconstructed energy. To suppress 
combinatorial background, the value of $\Delta E$ for each event is required to be within the
$\pm3\sigma_{\Delta E}$ around the peak of the $\Delta E$ distribution, where
$\sigma_{\Delta E}$ is the resolution of the distribution. 

\begin{figure*}[tb]
  \centering
  \includegraphics[width=1.0\textwidth]{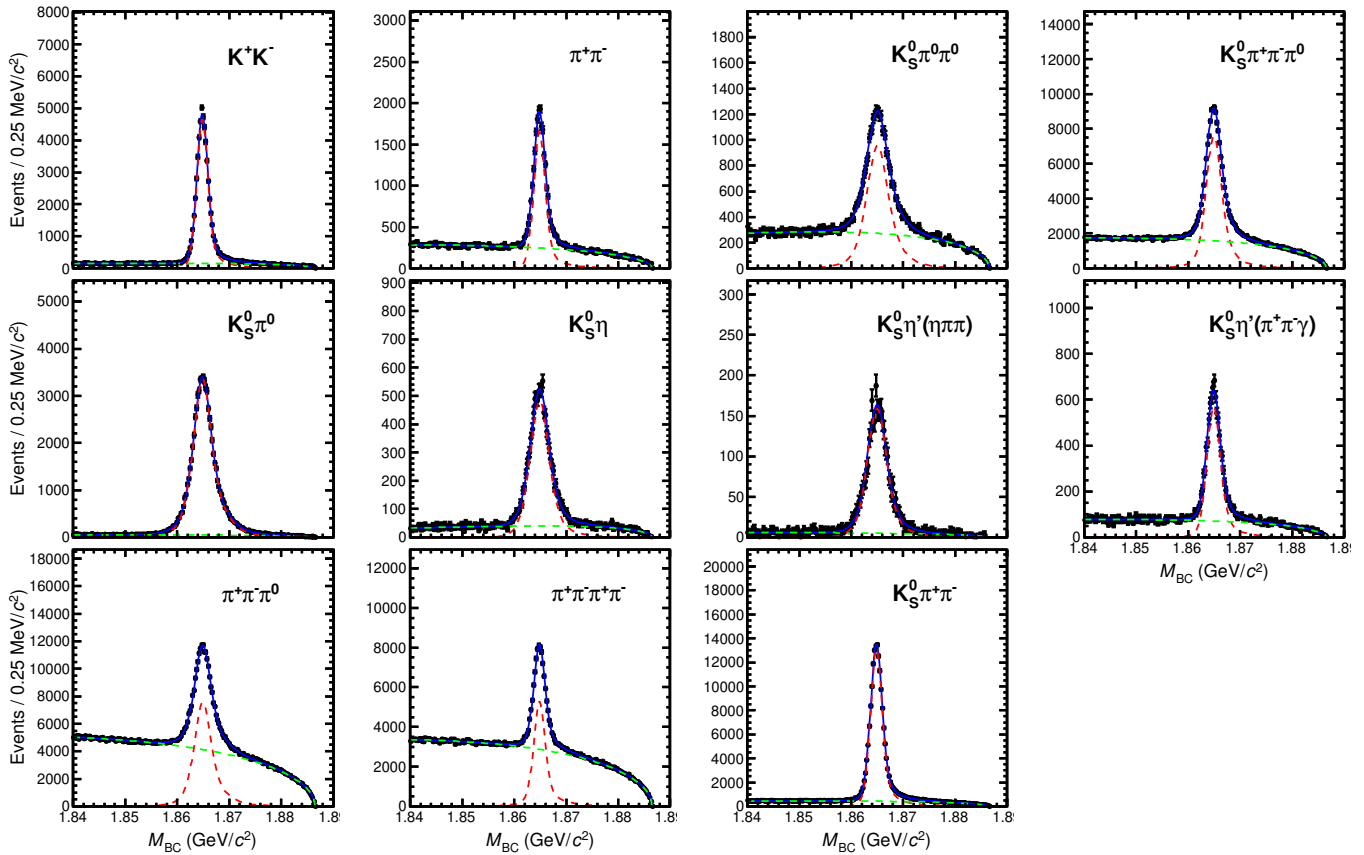}
  \caption{Fits to the $M_\mathrm{BC}$ distributions from the ST $D$
  candidates. The corresponding decay modes are denoted by the labels on the
plots. The black points with error bars represent data. The blue solid curves
are the fit results. The red and green dashed curves represent the signal and
background contributions of the fits, respectively.} \label{fig:STMBCFit}
\end{figure*}

\Cref{fig:STMBCFit} shows the $M_\mathrm{BC}$ distributions of the ST
$D$ candidates for the fully reconstructed tag modes. The ST yields for
the tag modes are obtained by an unbinned maximum likelihood fit to the
$M_\text{BC}$ distributions. In the fit, the signal component is described by the
MC-simulated shape. To account for the difference in 
resolutions between the MC simulation and data, the MC-simulated shape for each
mode is convolved with a Gaussian function with free parameters. The background component is modeled with an
ARGUS function~\cite{ARGUS:1990hfq}, where the slope parameter is a free parameter and
the endpoint is fixed to the beam energy~\cite{BESIII:2015zbz} in the
center-of-mass frame. The fit results are shown in \cref{fig:STMBCFit}.
In addition to the combinatorial background, there are also peaking backgrounds that
have similar final states as the signal and are included in the signal yield.
The contamination rate of the peaking background is estimated by analyzing the
inclusive MC sample. For the tag mode ${D\to K_S^0\pi^0}$~($K_S^0\pi^0\pi^0$), the
background level is 4.7\%~(0.4\%) dominated by ${D\rightarrow \pi^{+} \pi^{-}
\pi^{0}}$ ($\pi^{+} \pi^{-}\pi^{0}\pi^{0}$) decays. In the case of the ${D\to K_S^0 \eta^{\prime}
(\pi^+\pi^-\gamma)}$ tags,  the peaking background is dominated by ${D\to K_S^0\pi^+\pi^-\pi^0}$
decays with a contamination rate of 3.1\%. For 
$D\to \pi^{+} \pi^{-} \pi^{0}$ and ${D\to \pi^{+} \pi^{-} \pi^{+} \pi^{-}}$ tags, the
peaking backgrounds mainly originate from  $D\rightarrow K_{S}^{0} \pi^{0}$ and
${D\rightarrow K_{S}^{0} \pi^{+} \pi^{-}}$ decays, which have a contamination
rates of 5.2\% and 3.5\%, respectively. For the self-tag mode ${D\to K_S^0\pi^+\pi^-\pi^0}$,
the main peaking background is from the $D\rightarrow K_{S}^{0} \pi^{+}
\pi^- \gamma$ and $D\rightarrow K_S^0K_S^0\pi^0$ decays, constituting 1.1\% of
the yield. The ST yields after peaking background subtraction are
summarized in \cref{tab:dtwithsysErr}.

The $D\to K_L^0\pi^0$ and $D\to K_L^0\omega$ tag modes cannot be fully reconstructed.
Therefore, the effective ST yields are calculated
from \cref{eq:styields}. The effective ST efficiencies for the $D\to K_L^0\pi^0$ and $D\to K_L^0\omega$ tags,
which are required inputs for \cref{eq:styields}, are defined to be the ratios of the corresponding DT efficiencies to the
efficiency for the ST of the signal mode. The effective ST yields are also
summarized in \cref{tab:dtwithsysErr}.

\section{ Determination of the DT yield}\label{sec:dtyields}
\begin{figure}[t]
  \centering
  \includegraphics[width=0.47\textwidth]{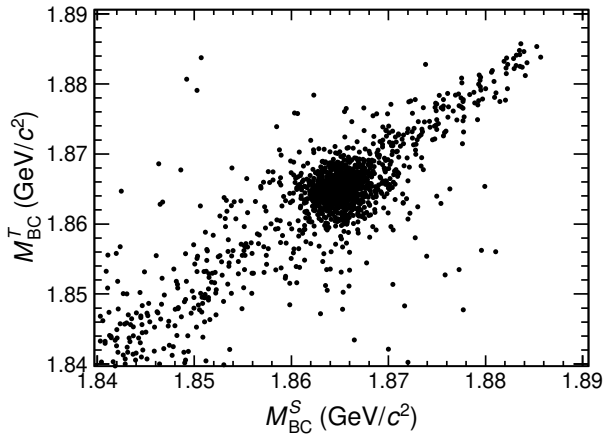}
  \caption{The distribution of $M_\text{BC}^{T}$ versus $M_\mathrm{BC}^{S}$
   for the DT candidates tagged by the $D\to \pi^+\pi^-\pi^0$
tag mode.}
  \label{fig:2DMBC}
\end{figure}

\begin{figure*}[bt]
  \centering
  \includegraphics[width=1.0\textwidth]{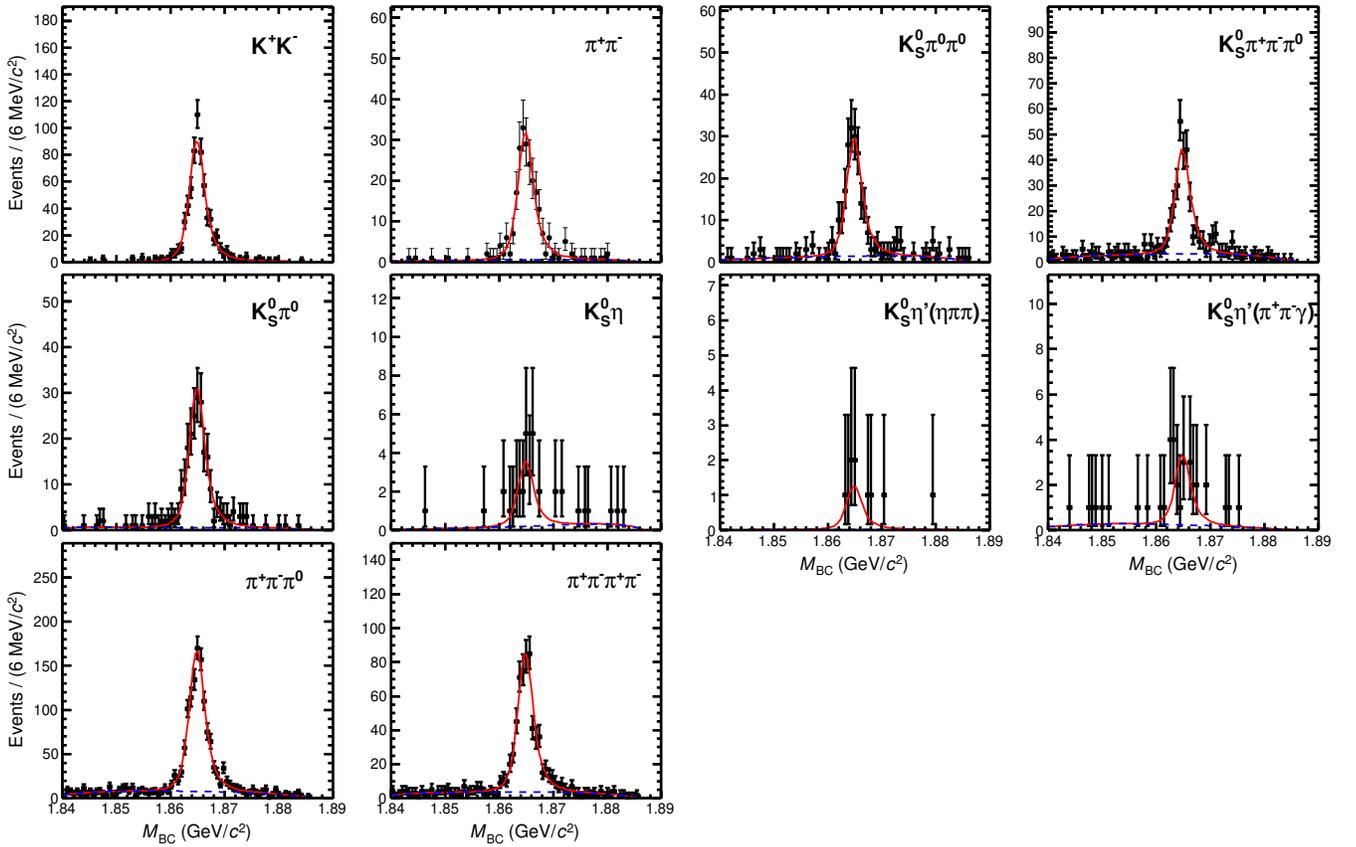}
  \caption{The projections of the 2D fits on the
    ${M}_{\mathrm{BC}}^{{S}}$ distribution. The black points represent the data. Overlaid
    is the fit projection in the continuous red line. The blue dashed line
  indicates the combinatorial component.}
  \label{fig:2dfitSD}
\end{figure*}

\begin{figure*}[bt]
  \centering
  \includegraphics[width=1.0\textwidth]{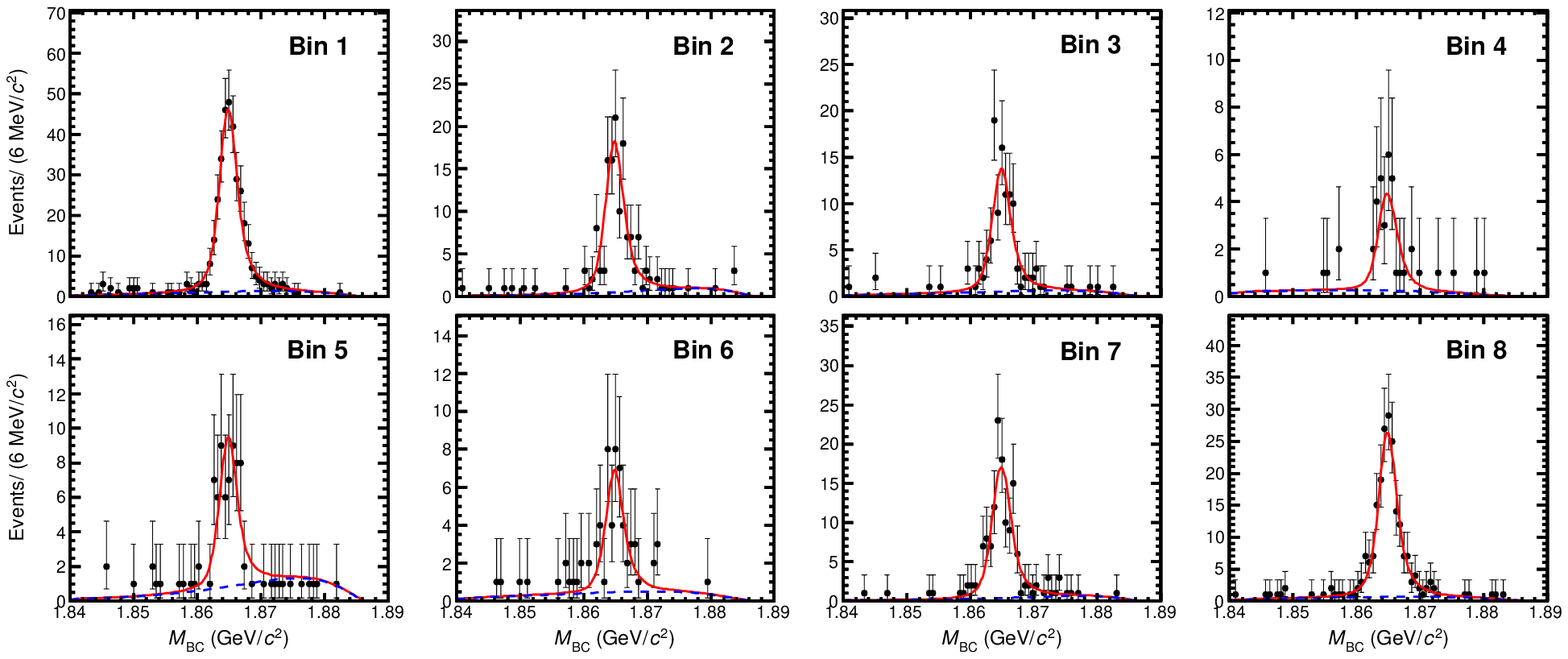}
  \caption{The projections of the 2D fits on the
    ${M}_{\mathrm{BC}}^{{S}}$ distribution in the eight pairs of bins for the
    $D\rightarrow K_S^0\pi^+\pi^-$ tag mode. The black points represent the data. Overlaid
    is the fit projection in the continuous red line. The blue dashed line
  indicates the combinatorial component.}
  \label{fig:2dfitSDKsPiPi}
\end{figure*}

The DT candidates for the fully reconstructed tag modes are isolated
with the beam-constrained masses for the signal and tag modes, denoted as
$M_\mathrm{BC}^S$ and $M_\mathrm{BC}^T$, respectively. In the case of
multiple combinations, the combination with the least $|M_\mathrm{BC}^S
+M_\mathrm{BC}^T -2M_\mathrm{PDG}^{D}|$ is retained for further analysis. The
$\Delta E$ variables for the signal and tag modes are
required to lie within the regions given in 
\cref{sec:styields}.  To determine the DT yield, a
two-dimensional~(2D) unbinned maximum likelihood fit is performed to the
$M_\mathrm{BC}^S$ versus $M_\mathrm{BC}^T$ distribution. An example of the
$M_\mathrm{BC}^T$ versus $M_\mathrm{BC}^S$ distribution for the $D\to \pi^+\pi^-\pi^0$ tag mode is shown in
\cref{fig:2DMBC}. The signal component
in the fit is described by a 2D MC-simulated
shape obtained from the signal MC sample smeared with a Gaussian function
in each dimension. The Gaussian function is introduced to account for the
difference in resolutions between the MC simulation and data.
The mass and width of the Gaussian function for each tag mode~(the signal
mode) are obtained from the one-dimensional fit in the ST-yield
determinations for the corresponding tag mode~(the signal mode). The
background component with the correctly reconstructed signal
mode~(tag mode) and incorrectly reconstructed tag mode~(signal mode) is
modeled by the product of the signal and background shapes from the fits for
the ST yield determinations of signal mode~(tag mode) and tag
mode~(signal mode), respectively. The parameters of the shapes are fixed at
the values
obtained from the corresponding one-dimensional fit. The background component where signal and
tag mode are
both reconstructed incorrectly is modeled by the product of the background
shapes from corresponding fits for the ST yield determinations. The
backgrounds involving swapped final-state particles from the two charm mesons
and continuum processes, corresponding to the diagonal band in \cref{fig:2DMBC},
are modeled by the product
of a Gaussian function and the ARGUS function rotated by
45\textdegree~\cite{BESIII:2020wnc}. The endpoint of the ARGUS function
is fixed at the beam energy in the $e^+e^-$ center-of-mass frame.
\Cref{fig:2dfitSD} shows the projections of 2D fits on the $M_\mathrm{BC}^S$
distributions for the fully reconstructed
tag modes. According to studies performed with the inclusive MC sample,
there are small contributions from peaking backgrounds for each mode. The dominant
components for the signal and tag modes are the same as those in the
determination of ST yields. The peaking background yields are determined by
analyzing the inclusive MC sample and are corrected for the quantum
correlation with \cref{eq:dtFp}. For the $D\to K_S^0\pi^+\pi^-$ tag mode, the DT yields in the eight 
pairs of bins are determined with the same method as described above. The fitted results are shown in \cref{fig:2dfitSDKsPiPi}.

\begin{figure}[t]
  \centering
  \subfigure{
    \includegraphics[width=0.47\textwidth]{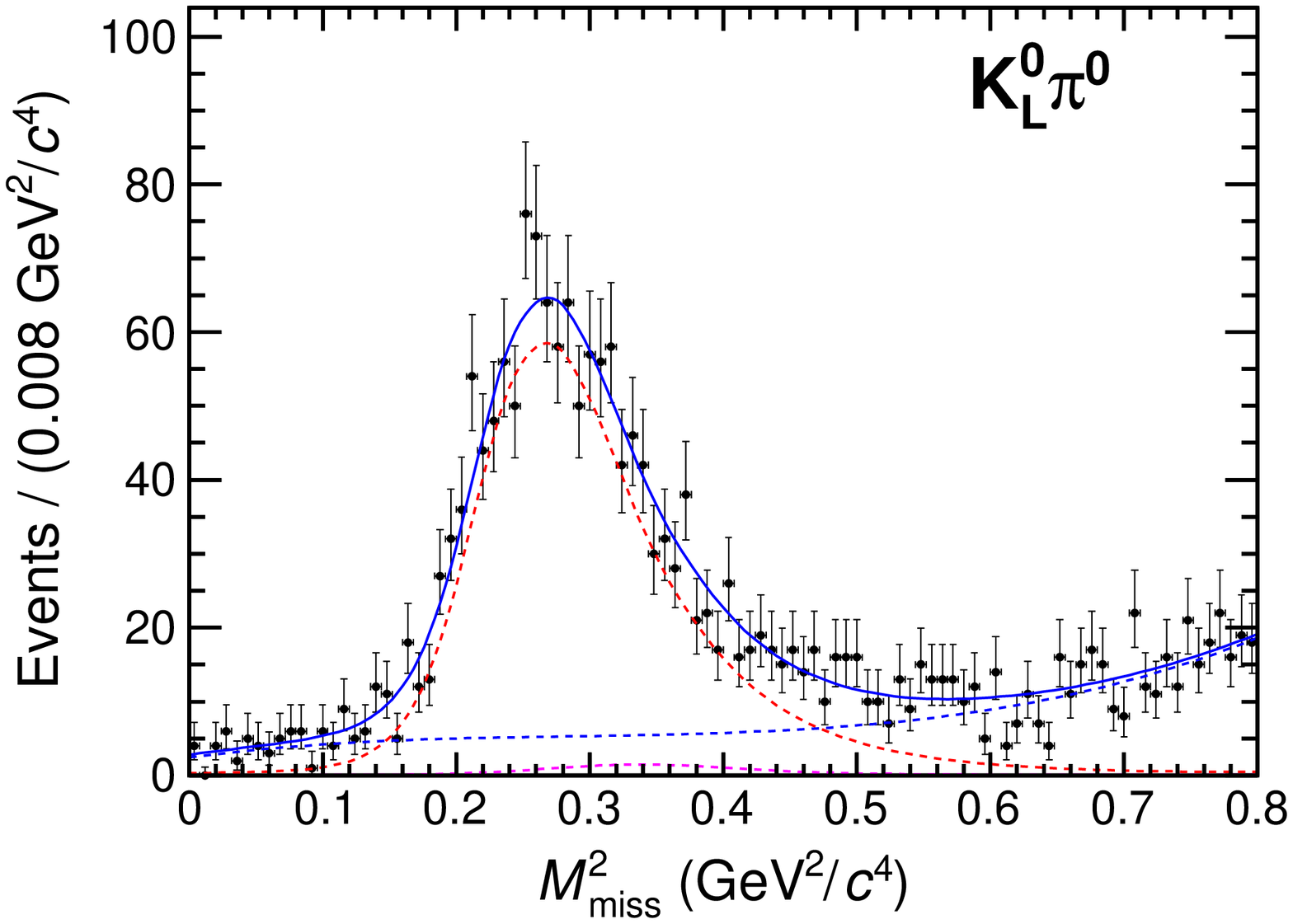}
  }
  \subfigure{
    \includegraphics[width=0.47\textwidth]{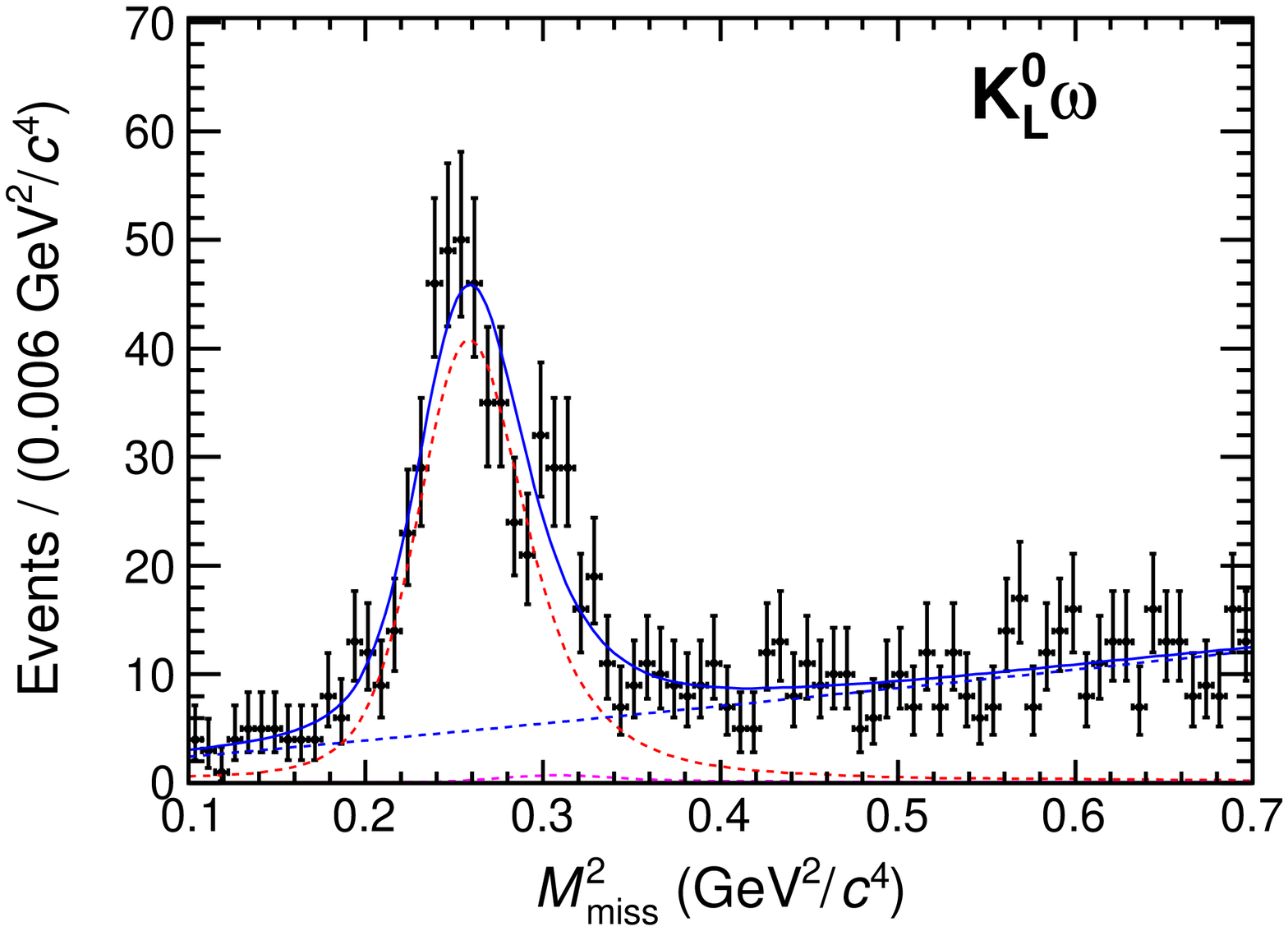}
  }
  \caption{Fits to the $M_\text{miss}^2$ distributions of the ${D\to K_L^0\pi^0}$~(top)  and
  $D\to K_L^0\omega$~(bottom) tag modes. The points with error bars represent data, the blue dashed
curves are the fitted combinatorial backgrounds, the dashed red and magenta lines show the
MC-simulated signal and peaking background shapes, respectively, and the blue solid curves
show the total fits.} \label{fig:mMiss2Fit}
\end{figure}

\begin{figure*}[bt]
  \centering
  \includegraphics[width=1.0\textwidth]{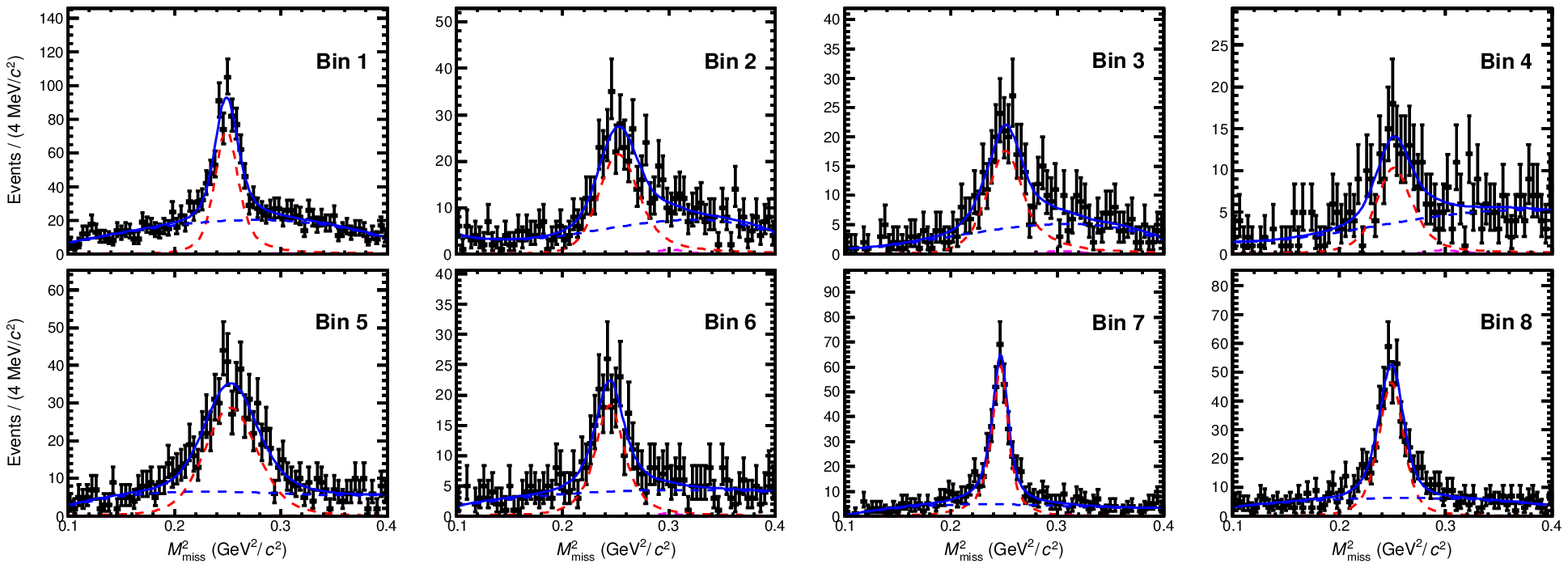}
  \caption{Fits to the $M_\text{miss}^2$ distributions in the eight pairs of
  bins of $D\to K_L^0\pi^+\pi^-$ tag mode. The points with error bars
represent data, the blue dashed curves are the fitted combinatorial
backgrounds, the dashed red and magenta lines show the MC-simulated signal and
peaking background shapes, respectively, and the blue solid curves show the
total fits.}
  \label{fig:mMissKLpipi}
\end{figure*}

The DT candidates tagged by the ${D\to K_L^0\pi^0}$ and ${D\to K_L^0\omega}$
tag modes cannot be
fully reconstructed. They are selected by the variable $M_\text{miss}^2$
defined as
\begin{equation}
  \label{eq:mMiss2}
  M_{\text {miss }}^2=\left( E_\mathrm{beam}-\sum_i E_i\right)^2-\left|\sum_i \vec{p}_i\right|^2,
\end{equation}
where $\sum_i E_i$ is the sum of the reconstructed energies of the tag mode and
$\sum_i \vec{p}_i$ is the sum of the reconstructed momenta of the signal mode and
tag mode. The distribution from correctly reconstructed
DT candidates peaks around the squared mass of the $K_L^0$ meson.
To suppress background, events with excess reconstructed 
charged tracks or $\pi^0$ candidates are rejected. The DT yields are
determined by fitting to the $M_\text{miss}^2$ distribution. In this fit, the
signal is described by an
MC-simulated shape, which is convolved with a Gaussian with free parameters.
The combinatorial
background is modeled with a second-order Chebyshev function. The
yield of peaking background for the $D\to K_S^0\pi^0$~($D\to K_S^0\omega$) decay is
estimated with the
inclusive MC sample and is corrected for quantum correlations.  The peaking
background in the $D\to K_L^0\pi^0$~($D\to K_S^0\omega$) tag mode originates from
$D\rightarrow\eta\pi^0$~($D\to \eta K_S^0\pi^0$) decays and is modeled by an
MC-simulated shape with its yield fixed according to the results from the study of the
inclusive MC sample. For the
$D\to K_L^0\omega$ tag mode, the non-resonant background from $D\to
K_{S,L}^0\pi^+\pi^-\pi^0$ is
estimated with the $\omega$ sideband events in the $M_{\pi^+\pi^-\pi^0}$
distribution. \Cref{fig:mMiss2Fit} shows the fit results for the two tag modes. 
The DT yields for the $D\to K_L^0\pi^+\pi^-$ tag mode in the eight pairs of bins
are determined with the same method. The corresponding fit results in the
eight pairs of bins are shown in \cref{fig:mMissKLpipi}.

\section{Systematic Uncertainties on the Yield determinations}\label{sec:sysErr}

\subsection{The ST yields}%
\label{subsec:sysStg}

The ST yields for the fully reconstructed tag modes are determined by
fitting the $M_\mathrm{BC}$ distributions after subtracting the peaking background
yields estimated from the inclusive MC sample. The uncertainty
associated with the fit is estimated by floating the endpoint, which is fixed
in the baseline fit, of the ARGUS function. The difference in the yield to that of the baseline fit 
is taken as the systematic uncertainty. The uncertainties for the different tag
modes lie in the range of [0.1, 0.3]\%. The uncertainties on the peaking
background yields are estimated by varying the quoted branching
fractions~\cite{Workman:2022ynf} by  $\pm1\sigma$ and range from 0.1\% to
0.5\%. 

For the $D\to K_L^0\pi^0$ and $D\to K_L^0\omega$ tag modes, the uncertainties for the effective
ST yields calculated as~\cref{eq:styields} are associated with
$N_{D \bar{D}}$ and the product of the branching fractions and efficiencies of the
two tag modes. The
uncertainty of $N_{D \bar{D}}$ which is 1.0\% has been estimated in the measurement of
$D\bar{D}$ cross section~\cite{BESIII:2018iev}. The uncertainties of the
products of the branching fractions and efficiencies for the $D\to K_L^0\pi^0$ and
$D\to K_L^0\omega$ tag modes have been estimated in the branching fraction measurements of
the two tag modes~\cite{BESIII:2022qkh, BESIII:2022xhe}. They are 3.1\% and
2.6\% for the $D\to K_L^0\pi^0$ and $D\to K_L^0\omega$ tag modes, respectively. These uncertainties are
propagated to the effective ST yields.

\subsection{The DT yields}~
\label{subsec:sysdt}

The DT yields for the fully reconstructed tag modes are determined by the
2D fits to the $M_\mathrm{BC}^S$ versus $M_\mathrm{BC}^T$ distributions
after subtracting the estimated peaking background after correcting for the effects of quantum correlations.
Since the fit strategies are the same for all tag modes, the
largest DT sample, which is that involving $D\to \pi^+\pi^-\pi^0$ tags, is adopted to estimate
the uncertainties introduced by the fits in order to minimize the effects of 
statistical fluctuations. For the uncertainty arising from the signal models,
the 2D MC-simulated shape without a smearing Gaussian resolution function is taken as the
alternative model. The change of the signal yield, 0.34\%, is taken as the
uncertainty. For the background shapes, the
fixed parameters of the ARGUS functions are changed to free parameters in the fit. The change in the signal yield, which is 0.13\%,
is assigned as the corresponding uncertainty. The uncertainties due to
the peaking background subtraction with the inclusive MC sample are estimated with the
uncertainties of the corresponding branching fractions~\cite{ Workman:2022ynf} and $F_+$ for
quantum correlations~\cite{Resmi:2017fuo, Malde:2015mha, BESIII:2022wqs,
BESIII:2020khq}. The estimated uncertainties are in the range of [0.1, 1.0]\%. The propagated systematic uncertainties for the tag modes are
listed in \cref{tab:dtwithsysErr}.

For the $D\to K_L^0\pi^0$ and $D\to K_L^0\omega$ tag modes, the yields are determined from the
fits to the $M_\text{miss}^2$ distribution after subtracting the peaking background yield
estimated from the
inclusive MC sample and data studies. The systematic uncertainties
from the fits to the $M_\mathrm{miss}^2$ distributions have several components.  To assess that uncertainty coming
from the background shape, the second-order Chebyshev function is replaced by a third-order Chebyshev
function. The resulting differences in the yields, 1.1\% for the $D\to K_L^0\pi^0$ tag mode and
0.17\% for the $D\to K_L^0\omega$ tag mode, are assigned as the
uncertainties from the fit procedure. The uncertainties from the peaking
background subtractions are estimated by varying the
assumed branching fractions~\cite{ Workman:2022ynf}  and $F_+$ for the quantum correlation corrections~\cite{Resmi:2017fuo, Malde:2015mha, BESIII:2022wqs,
BESIII:2020khq} by  $\pm1\sigma$. For the
$D\to K_L^0\omega$ tag mode, the contribution of the peaking background from non-resonant
$D\to K_{S,L}^0 \pi^+ \pi^- \pi^0$ final state is estimated by fitting the
$M_\mathrm{miss}^2$ distribution from the $\omega$ sidebands. The uncertainty on
the fitted peaking background yield is taken as the uncertainty from the sample
size of sideband events. The sidebands are altered to estimate the uncertainty
due to the choice of sidebands. The estimated uncertainties from the peaking
background subtractions are 0.32\% and 2.1\% for the $D\to K_L^0\pi^0$ and
$D\to K_L^0\omega$ tag modes, respectively. The combined systematic
uncertainty of the DT yield for each tag mode is summarized in
\cref{tab:dtwithsysErr}.

\begin{table}[b]
  \centering
  \caption{The DT and ST yields~($N_\mathrm{DT}$ and
    $N_\mathrm{ST}$) for each tag mode. For the fully reconstructed tag
    modes, the first uncertainties are statistical and the second are systematic.}
\label{tab:dtwithsysErr}
\begin{ruledtabular}
  \begin{tabular}{ccc}
    Mode & $N_\text{DT}$ & $N_\text{ST}$ \\
    \hline
    $ K^+K^-          $ & $\phantom{0}602.5\pm\phantom{0}25.2\pm\phantom{0}3.1$ & $  56088.8  \pm   254.7\pm\phantom{0}39.3 $  \\
    $ \pi^+\pi^-      $ & $\phantom{0}215.4\pm\phantom{0}15.6\pm\phantom{0}1.1$ & $  20601.9  \pm   179.1\pm\phantom{0}33.0 $  \\
    $ K_S^0\pi^0\pi^0 $ & $\phantom{0}180.5\pm\phantom{0}14.9\pm\phantom{0}1.2$ & $  21871.9  \pm   212.0\pm          100.6 $  \\
    $ K_L^0\pi^0      $ & $          1209.4\pm          102.0\pm          14.2$ & $           120901.3\pm 3994.0 $                \\
    $ K_L^0\omega     $ & $\phantom{0}402.1\pm\phantom{0}29.5\pm\phantom{0}9.4$ & $ \phantom{0}48400.5\pm 1373.7  $                 \\
    \hline
    $ K_S^0\pi^0                           $ & $                      207.5\pm          15.6 \pm 1.3$ & $           71046.2 \pm            284.4 \pm          56.8  $  \\
    $ K_S^0\eta                            $ & $            \phantom{0}23.3\pm\phantom{0}5.3 \pm 0.1$ & $ \phantom{0}9647.3 \pm            114.6 \pm\phantom{0}3.9  $  \\
    $ K_S^0\eta^{\prime}_{\pi^+\pi^-\eta}  $ & $ \phantom{0}\phantom{0}5.3 \pm\phantom{0}2.5 \pm 0.0$ & $ \phantom{0}3250.5 \pm  \phantom{0}62.4 \pm\phantom{0}8.8  $  \\
    $ K_S^0\eta^{\prime}_{\pi^+\pi^-\gamma}$ & $            \phantom{0}21.4\pm\phantom{0}5.2 \pm 0.1$ & $ \phantom{0}7954.4 \pm            107.2 \pm          29.4  $  \\
    \hline
    $ K_S^0\pi^+\pi^-\pi^0 $      & $257.8 \pm 18.3 \pm 1.8$ & $ 122688.0  \pm 455.0   \pm 147.2 $ \\
    $ \pi^+\pi^-\pi^0      $      & $990.8 \pm 35.2 \pm 6.8$ & $ 113407.9 \pm 591.0  \pm 139.1  $ \\
    $ \pi^+\pi^-\pi^+\pi^-      $ & $503.1 \pm 27.6 \pm 5.7$ & $ 68274.9  \pm 429.6  \pm 84.7 $    \\
  \end{tabular}
\end{ruledtabular}
\end{table}

\section{The \texorpdfstring{$F_+$}{Fp} measurement}%
\label{sec:FpMea}

\subsection{Measurement with the \texorpdfstring{$C\!P$}{CP}-tag modes}%
\label{subsec:FpCP}

\begin{figure}[tb]
  \centering
  \subfigure{
    \includegraphics[width=0.47\textwidth]{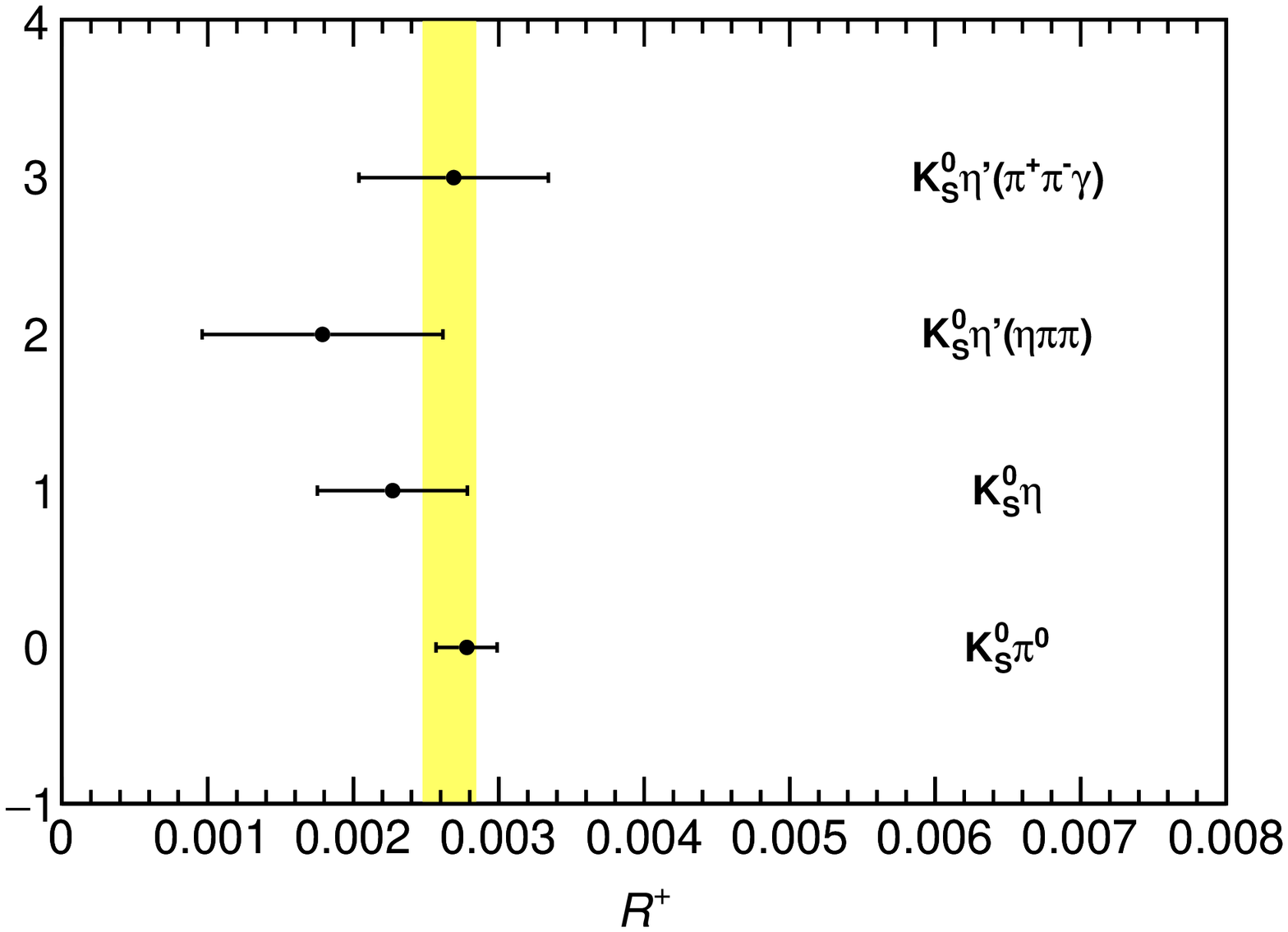}
  }\\
  \subfigure{
    \includegraphics[width=0.47\textwidth]{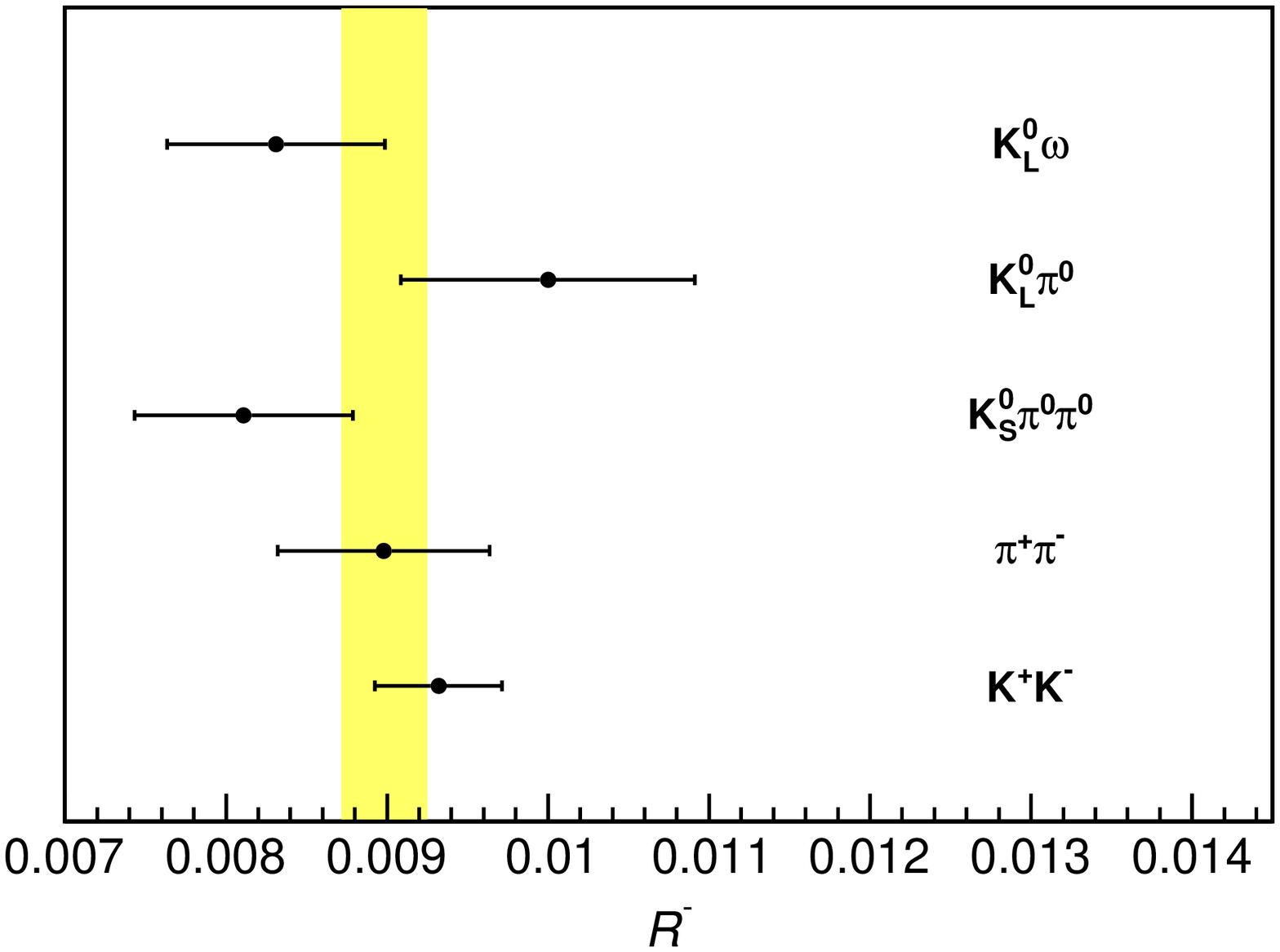}
  }
  \caption{The $R^+$values~(top) for the $C\!P$-odd tag modes and the
     $R^-$ values~(bottom) for the $C\!P$-even tag modes. The horizontal error
   bars show the total uncertainty for each measurement. The yellow bands show the
 fitted values with uncertainties.}\label{fig:npnm}
\end{figure}

The expected ratio of the DT yield to the corresponding ST yield is calculated from 
\cref{eq:norDT_CP}. Implicity in this expression is the assumption of $\varepsilon(S\mid T) = \varepsilon(S)
\cdot\varepsilon(T)$. However, studies of the signal MC samples indicate that this assumption is not always true.  Therefore, a correction factor of
$\varepsilon(S)\cdot\varepsilon(T)/\varepsilon(S|T)$ is applied to the measurement of $R^\pm$
for each tag mode. \Cref{fig:npnm} shows the measured $R^\pm$ values for each tag mode after applying this correction.
The mean values of 
$R^+$ and $R^-$ are determined by least $\chi^2$ fits. The $\chi^2_\pm$
for $R^\pm$ in the fit is constructed as follows:
\begin{equation} 
  \label{eq:chi2NpNm}
  \chi^2_\pm = \sum_{ij} (R_i^{\pm} -
\langle R^\pm \rangle) (R_j^\pm - \langle R^\pm \rangle) (V^\pm )^{- 1}_{ij},
\end{equation} 
where $\left\langle R^\pm\right\rangle$ is the mean value of $R^\pm$,
$R^\pm_i$~($R^\pm_j$) is the ratio of the DT yield to the corresponding ST
yield of the $i\text{-th}$~($j\text{-th}$) tag mode, and $V^\pm_{ij}$ is the
covariance between modes $i$ and $j$.
The
measured values of $R^\pm$ for the different tag modes are independent, except for the $D\to K_L^0\pi^0$ and
$D\to K_L^0\omega$ tags, where there is a correlation coefficient of 0.02 introduced by the common use of $N_{D \bar{D}}$. The fitted
result for $\left\langle R^{\pm}\right\rangle$ is shown as the yellow bands in
\cref{fig:npnm}.  These results from the $C\!P$-tag modes lead to a value of $F_+ = 0.229\pm 0.013$, where the
uncertainty includes both statistical and
systematic contributions.  Re-performing the fit with only the statistical uncertainties included on the inputs 
allows the statistical and systematic contributions on the fit uncertainty to be isolated,
with the statistical uncertainty found to be $0.013$ and the systematic uncertainty to be $0.001$.

It is necessary to apply a correction to this result for $F_+$ to account for the 
fact that the signal efficiency is in principle different for DTs involving $C\!P$-even
and $C\!P$-odd tags. This is because the distribution over phase space of final-state
particles will be different for decays of the signal mode when it is tagged as 
$C\!P$-odd or $C\!P$-even. For example, the intermediate process $D\rightarrow K_S^0\eta$ exists in 
signal decays tagged by the $C\!P$-odd eigenstates, but not for $C\!P$-even tags.
Comparison of DT events containing $C\!P$-even and $C\!P$-odd tags show the 
momentum and $\cos \theta$ distributions of the signal decay are very similar, 
apart from that of the $K_S^0$ momentum.  Studies of these $K_S^0$ momenta distributions,
together with the known variation in reconstruction efficiency with $K_S^0$ momentum~\cite{BESIII:2017ylw},
is used to determine that the ratio of the efficiency of the signal mode tagged by $C\!P$-even eigenstates 
to that of the efficiency of the signal mode tagged by $C\!P$-odd eigenstates
is $1.008\pm
0.06$. Applying this ratio as a correction leads to the result from
$C\!P$-tagged events of $F_+=0.229\pm 0.013 \pm 0.002$.
An additional systematic component for the potential difference in the
efficiencies has been included, which is estimated by the
difference between this central value and the one 
obtained from the corrected fit.
This result is consistent with that obtained from CLEO-c
data~\cite{Resmi:2017fuo} with $C\!P$-tag modes and is a factor 1.6 more precise.

\subsection{Measurements with the quasi-\texorpdfstring{$C\!P$-tag}{CP} and self-tag modes}%
\label{subsec:fpMixedCP}

Using \cref{eq:FpMixedCP}, $F_+$ is determined with the quasi-$C\!P$-tag modes
$D\to \pi^+\pi^-\pi^0$ and $D\to \pi^+\pi^-\pi^+\pi^-$, where the $R^{T}$ is
also corrected with the factor \mbox{$\varepsilon(S)\cdot\varepsilon(T)/\varepsilon(S|T)$} for the same reason mentioned in \cref{subsec:FpCP}. The $C\!P$-even fractions of these modes, denoted as
$F_+^{\pi^+\pi^-\pi^0}$ and $F_+^{\pi^+\pi^-\pi^+\pi^-}$, are taken from
Refs.~\cite{Malde:2015mha,BESIII:2022wqs}. The value of $R^+$ is taken from the
measurement with the $C\!P$-tag modes described in \cref{subsec:FpCP}. The ratios of
the DT yields to the corresponding ST yields of the $D\to
\pi^+\pi^-\pi^0$ and $D\to \pi^+\pi^-\pi^+\pi^-$ tag modes are
calculated with the corresponding ST and DT yields listed in
\cref{tab:dtwithsysErr}. After propagating the uncertainties from the
input parameters, $F_+$ is determined to be $0.227\pm 0.014 \pm 0.003$ with $D\to
\pi^+\pi^-\pi^0$ tags
and $0.227\pm 0.016\pm 0.003$ with  $D\to \pi^+\pi^-\pi^+\pi^-$ tags. Here the systematic uncertainties are
assigned in the same way as for the measurement with the $C\!P$-tag modes.

The self-tag yield is also sensitive to $F_{+}$, as shown in \cref{eq:fpks3pi}. The ratio of the DT yield for the
self-tag mode to the corresponding ST yield is determined
with the corresponding yields listed in \cref{tab:dtwithsysErr} and is
corrected with the factor \mbox{$\varepsilon(S)\cdot\varepsilon(T)/\varepsilon(S|T)$}. The value of $R^-$ is 
taken from the measurement with the CP-tag modes. The $C\!P$-even fraction measured from the
self-tag mode is $F_+ = 0.244\pm 0.019\pm 0.002$. Here the systematic uncertainty
is assigned using the same method as for the measurement with the $C\!P$-tag modes.

\subsection{Measurement with the  \texorpdfstring{$D\to K_S^0\pi^+\pi^-$}{kspippim} and
\texorpdfstring{$D\to K_L^0\pi^+\pi^-$}{klpippim} tag modes}%
\label{subsec:fpKsKlpippim}

\begin{figure}[t]
  \centering
  \includegraphics[width=0.47\textwidth]{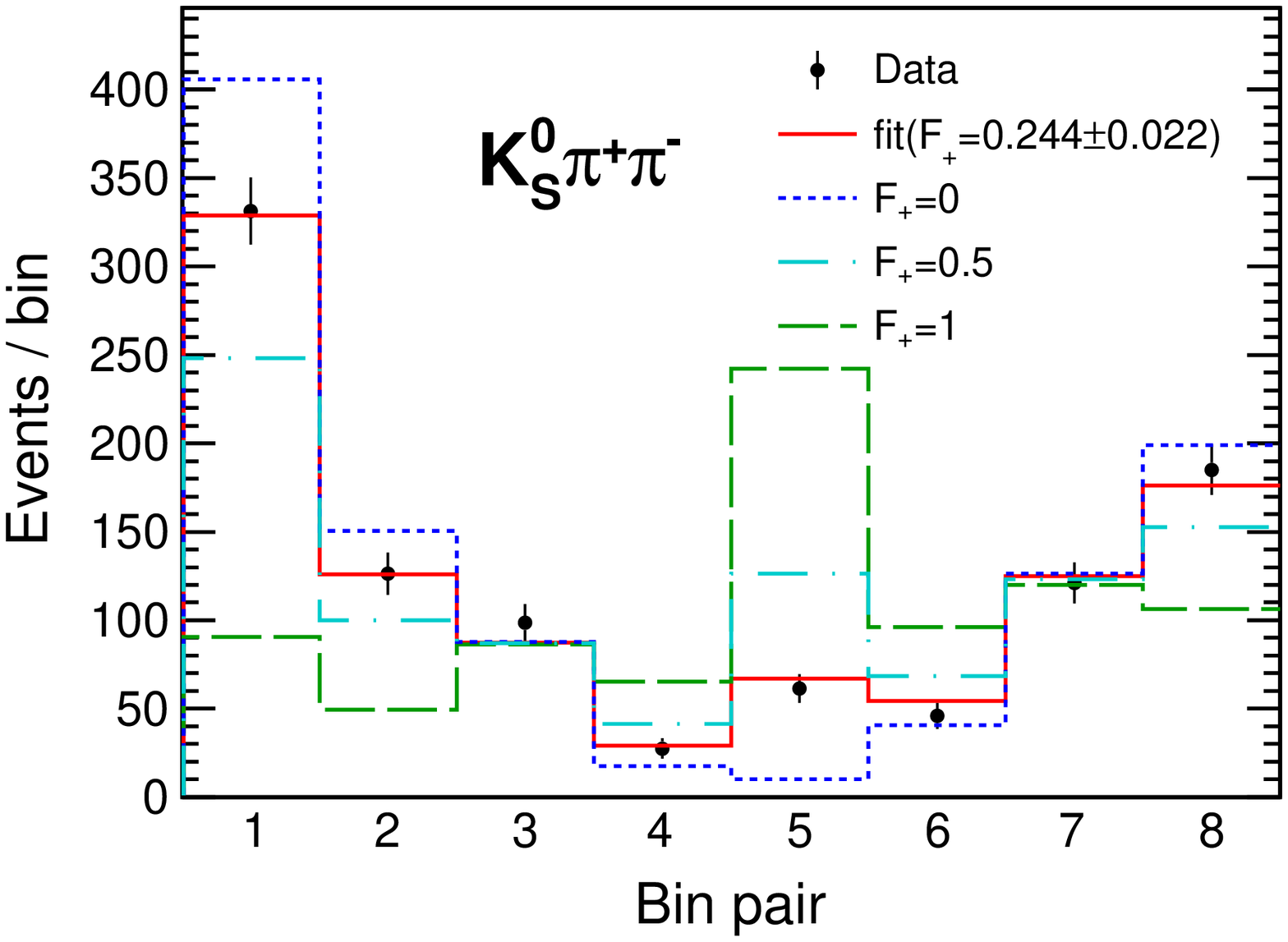}
  \\
  \includegraphics[width=0.47\textwidth]{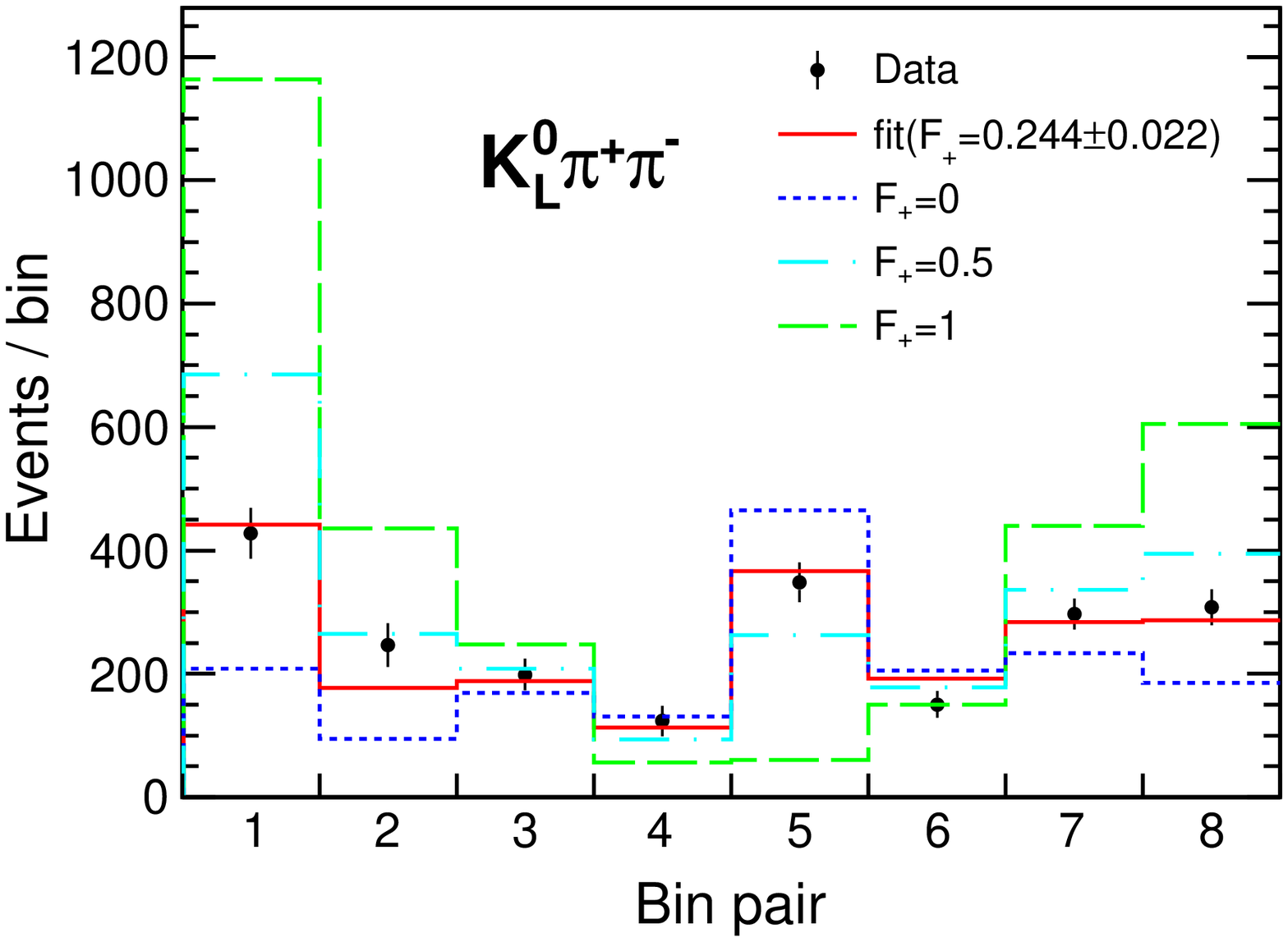}
  \caption{Predicted and measured yields for the ${D\to K_{S}^{0} \pi^{+} \pi^{-}
    }$~(top) and $D\to K_L^0\pi^+\pi^-$~(bottom) tag modes in each pair of bins. The black points with
error bars show the measured values. The red lines show the predicted values
from the fit, the dashed blue lines correspond to $F_{+}=0$, the dashed-dotted
cyan lines are for $F_{+}=0.5$ corresponding to no quantum correlation, and the
dashed-dotted green lines present the expected yields with $F_+=1$.} \label{fig:FpfpKsKlpipi}
\end{figure}

The measurement of $F_+$ with the
$D\to K_S^0\pi^+\pi^-$ and $D\to K_L^0\pi^+\pi^-$ tag modes is performed with the
measurements of the populations in the eight bin-pairs for the two tag
modes~\cite{Malde:2015mha}. The measured DT yields for the
$D\to K_S^0\pi^+\pi^-$ and $D\to K_L^0\pi^+\pi^-$ tag modes after subtracting peaking background
in the eight bin pairs are shown in \cref{fig:FpfpKsKlpipi},
which are determined with the same strategies described in \cref{sec:dtyields}.
Eq.~\ref{eq:fpKspipi} and~\ref{eq:fpKlpipi} are modified to account for migration effects and variations in bin-to-bin efficiencies, such that the expected DT yield in the $i$-th
bin pair as a function of $F_+$ is given by
\begin{equation}
  \label{eq:dtyieldKsPipiMig}
  \begin{aligned}
    M_i  = & h \sum_{j = 1}^8
    \varepsilon_{ij} \left[K_j + K_{- j} 
      - 2 c_j  \sqrt{K_j K_{- j}} (2 F_+ - 1) \right],
  \end{aligned}
\end{equation}
for $D\to K_S^0\pi^+\pi^-$  tags, and
\begin{equation}
  \label{eq:dtyieldKlPipiMig}
  \begin{aligned}
    M_i^\prime =  h'  \sum_{j = 1}^8 \varepsilon_{ij}'  \left[K_j^\prime + K_{- j}^\prime + 2 c_j^\prime \sqrt{K_j^\prime K_{- j}^\prime } (2 F_+ - 1) \right],
  \end{aligned}
\end{equation}
for  $D\to K_L^0\pi^+\pi^-$ tags, 
where $\varepsilon_{ij}$~($\varepsilon_{ij}^\prime$) is the migration matrix,
determined from MC simulation, 
describing the efficiency for an event produced in the $j$-th bin and
reconstructed in the $i$-th bin. To determine the value of $F_+$, a log-likelihood fit is
performed. The likelihood is given by
\begin{equation}
  \label{eq:logDT}
  \begin{aligned}
    - 2 \log \mathcal{L}= & - 2 \sum_{i = 1}^8 \ln G (N_i^{\mathrm{obs}},
    N_i^{\exp} (F_+), \sigma_{N_i^{\mathrm{obs}}})_{K_S^0}\\
    & - 2 \sum_{i = 1}^8 \ln G (N_i^{\mathrm{obs}}, N_i^{\mathrm{exp}} (F_+),
    \sigma_{N_i^{\mathrm{obs}}})_{K_L^0}\\
    & + \sum_{i = - 8}^8 \left( \frac{K_i -
    K_i^{\mathrm{inp}}}{\sigma_{K_i^{\mathrm{inp}}}} \right)_{K_S^0}^2\\
    & + \sum_{i = - 8}^8 \left( \frac{K_i -
    K_i^{\mathrm{inp}}}{\sigma_{K_i^{\mathrm{inp}}}} \right)_{K_L^0}^2\\
    & + \sum_{i = 1}^{16} \sum_{j = 1}^{16} (c_i - c_i^{\mathrm{inp}})  (c_j -
    c_j^{\mathrm{inp}}) (V^{- 1})_{ij}\\
    & + \sum_{i = 1}^{16} \sum_{j = 1}^{16} \left(
      \frac{\varepsilon_{ij} - \varepsilon_{i
      j}^{\mathrm{inp}}}{\sigma_{\varepsilon_{i
  j}^{\mathrm{inp}}}} \right)^2_{K_S^0}\\
    & + \sum_{i = 1}^{16} \sum_{j = 1}^{16} \left(
      \frac{\varepsilon_{ij} - \varepsilon_{i
      j}^{\mathrm{inp}}}{\sigma_{\varepsilon_{i
  j}^{\mathrm{inp}}}} \right)_{K_L^0}^2,
  \end{aligned}
\end{equation}
where $N^\mathrm{exp}_i$ is the expected yield in the $i$-th bin pair as a
function of $F_+$, $N^i_\mathrm{obs}$ is the observed yield with peaking
background subtracted in the $i$-th bin, $\sigma_{N^i_\mathrm{obs}}$ is the
uncertainty of the
observed yield in the $i$-th bin, $K_i$ and $\sigma_{K_i}$ are the
flavor-tagged fraction in bin $i$ and its uncertainty, respectively and $c_i$ is the strong-phase
parameter of the tag mode in bin $i$ with covariance matrix $V$.
The $K_i$ and $c_i$ parameters are fit parameters, but constrained through Gaussian functions. The means~( $K_i^\mathrm{inp}$ and
$c_i^\mathrm{inp}$) and covariances~($\sigma_{K_i^\mathrm{inp}}$ and $V_{ij}$)
of the Gaussian functions for $K_i$ and $c_i$ are taken from the combined results
from the BESIII and CLEO Collaborations~\cite{BESIII:2020khq}. The elements of the migration matrix are also fit parameters,
but included with  $\chi^2$ constraints, with mean $\varepsilon^\mathrm{inp}_{ij}$ and width
$\sigma_{\varepsilon^\mathrm{inp}_{ij}}$, where the uncertainties arise from from the finite size of MC sample. 
The fit is performed twice, once with the full uncertainties included, and then with only the statistical contributions.
From these fits, it is found that $F_+$ is $0.244\pm 0.021\pm0.006$, where the first uncertainty is statistical, and the second is systematic. 
Figure~\ref{fig:FpfpKsKlpipi} shows the DT yield in each bin, together with the expected yields with the fitted value of $F_+$, and the expected yields with other
values of $F_+$. 
Individual fits performed with each tag mode separately return compatible results, 
with $0.211\pm 0.029$ for  $D\to K_S^0\pi^+\pi^-$ tags and $0.290\pm
0.037$  for $D\to K_L^0\pi^+\pi^-$ tags, where the combined statistical and systematic uncertainties are given.

\subsection{Combination of results}
\label{subsec:fpCombi}

\Cref{tab:Fpsum} summarizes the results of $F_+$ determined with different
tag modes, which are seen to be consistent with each other. A least-squared fit is
performed to combine the results of $F_+$, taking the uncertainties and
correlations between the results into account. The correlations are introduced
because of the common use of $\left\langle R^+\right\rangle$ and $\left\langle R^-
\right\rangle$ in~\cref{eq:fpCP,eq:FpMixedCP,eq:fpks3pi}. The correlation coefficients
between the correlated tag modes are summarized in \cref{tab:CorreCoe}. The
combined result from all tags $F_+$ is $0.235\pm 0.010\pm 0.002$, which is
consistent with the result $0.238\pm0.012\pm0.012$~\cite{Resmi:2017fuo} based
on the CLEO-c data, but is a factor 1.7 times more precise. The
result is also compatible with the value $F_+= 0.226\pm0.020$ deduced from
the strong-phase parameters $c_i$, also determined with CLEO-c data~\cite{Resmi:2017fuo}.

\begin{table}[tb]
  \centering
  \caption{Results of $F_+$ from different tag modes and the combination
    of these results, where the first
  uncertainties are statistical and the second are systematic.}
  \label{tab:Fpsum}
  \begin{ruledtabular}
    \begin{tabular}{cc}
      Method                                       & $F_{+}$ \\
      \hline
      $C\!P$-tag modes                & $0.229 \pm 0.013 \pm 0.002$ \\
      $\pi^+\pi^-\pi^0$ tag mode      & $0.227 \pm 0.014 \pm 0.003$ \\
      $\pi^+\pi^-\pi^+\pi^-$ tag mode & $0.227 \pm 0.016 \pm 0.003$ \\
      Self-tag modes                  & $0.244 \pm 0.019 \pm 0.002$ \\
      $K_{S,L}^{0} \pi^{+} \pi^{-}$   & $0.244 \pm 0.021 \pm 0.006$ \\
      \hline
      Combined & $0.235\pm 0.010\pm 0.002$\\
    \end{tabular}
  \end{ruledtabular}
\end{table}

\begin{table}[tb]
  \centering
  \caption{Correlation coefficients between the results measured by
  different types of tag modes.}
  \label{tab:CorreCoe}
  \begin{ruledtabular}
    \begin{tabular}{cc}
      Tag mode & Correlation coefficient \\
      \hline
      $C\!P$ tag versus  $\pi^+\pi^-\pi^0$                 & 0.80\\
      $C\!P$ tag versus  $\pi^+\pi^-\pi^+\pi^-$            & 0.69\\
      $\pi^+\pi^-\pi^0$ versus  $\pi^+\pi^-\pi^+\pi^-$ & 0.66\\
      $C\!P$ tag versus self-tag             & 0.16\\
    \end{tabular}
  \end{ruledtabular}
\end{table}

\section{Summary}
\label{sec:summary}

The $C\!P$-even fraction $F_+$ of the $D^0\rightarrow K_{S}^{0}\pi^+\pi^-\pi^0$
decay has been measured by analyzing $2.93\mathrm{~fb}^{-1}$ of data
collected at $\sqrt{s}=3.773 \mathrm{~GeV}$ with the BESIII detector. The
measurement is performed with five categories of tag mode listed in \Cref{tab:Fpsum}, which give a 
consistent set of results. The combined result is $F_+=0.235\pm 0.010\pm 0.002$, where the first
uncertainty is statistical and the second is systematic. This result is
consistent with that obtained from CLEO-c
data~\cite{Resmi:2017fuo}, but is a factor 1.7 times more precise. The measured $F_+$
is an important input for the measurement of the unitarity triangle angle
$\gamma$ in $B\to D K$,  $D\to K_S^0\pi^+\pi^-\pi^0$ decays. Currently, the measurement
is dominated by statistical uncertainty. A future larger data
sample~\cite{Wilkinson:2021tby} allows us to improve the precision
significantly.

\acknowledgments
\input{acknowledgement_2023-01-18.tex}
\bibliography{ref}

\end{document}

%% file: authorlist_2023-01-18.tex
M.~Ablikim$^{1}$, M.~N.~Achasov$^{13,b}$, P.~Adlarson$^{75}$, X.~C.~Ai$^{81}$, R.~Aliberti$^{36}$, A.~Amoroso$^{74A,74C}$, M.~R.~An$^{40}$, Q.~An$^{71,58}$, Y.~Bai$^{57}$, O.~Bakina$^{37}$
% ...
\\
\vspace{0.2cm}
(BESIII Collaboration)\\
\vspace{0.2cm} {\it
$^{1}$ Institute of High Energy Physics, Beijing 100049, People's Republic of China\\
% ...
$^{13}$ G.I. Budker Institute of Nuclear Physics SB RAS (BINP), Novosibirsk 630090, Russia\\
% ...
$^{36}$ Johannes Gutenberg University of Mainz, Johann-Joachim-Becher-Weg 45, D-55099 Mainz, Germany\\
$^{37}$ Joint Institute for Nuclear Research, 141980 Dubna, Moscow region, Russia\\
% ...
$^{40}$ Liaoning Normal University, Dalian 116029, People's Republic of China\\
% ...
$^{57}$ Southeast University, Nanjing 211100, People's Republic of China\\
$^{58}$ State Key Laboratory of Particle Detection and Electronics, Beijing 100049, Hefei 230026, People's Republic of China\\
% ...
$^{71}$ University of Science and Technology of China, Hefei 230026, People's Republic of China\\
% ...
$^{74}$ University of Turin and INFN, (A)University of Turin, I-10125, Turin, Italy; (B)University of Eastern Piedmont, I-15121, Alessandria, Italy; (C)INFN, I-10125, Turin, Italy\\
$^{75}$ Uppsala University, Box 516, SE-75120 Uppsala, Sweden\\
% ...
$^{81}$ Zhengzhou University, Zhengzhou 450001, People's Republic of China\\

\vspace{0.2cm}
% ...
$^{b}$ Also at the Novosibirsk State University, Novosibirsk, 630090, Russia\\
% ...

}

%% file: acknowledgement_2023-01-18.tex
\textbf{Acknowledgement}

The BESIII Collaboration thanks the staff of BEPCII and the IHEP computing
center for their strong support. This work is supported in part by National
Key R\&D Program of China under Contracts Nos. 2020YFA0406400, 2020YFA0406300;
National Natural Science Foundation of China (NSFC) under Contracts Nos.
10975093 11635010, 11735014, 11835012, 11935015, 11935016, 11935018,
11961141012, 12022510, 12025502, 12035009, 12035013, 12061131003, 12192260,
12192261, 12192262, 12192263, 12192264, 12192265, 12221005, 12225509,
12235017; the Chinese Academy of Sciences (CAS) Large-Scale Scientific
Facility Program; the CAS Center for Excellence in Particle Physics (CCEPP);
CAS Key Research Program of Frontier Sciences under Contracts Nos.
QYZDJ-SSW-SLH003, QYZDJ-SSW-SLH040; 100 Talents Program of CAS; The Institute
of Nuclear and Particle Physics (INPAC) and Shanghai Key Laboratory for
Particle Physics and Cosmology; ERC under Contract No. 758462; European
Union's Horizon 2020 research and innovation programme under Marie
Sklodowska-Curie grant agreement under Contract No. 894790; German Research
Foundation DFG under Contracts Nos. 443159800, 455635585, Collaborative
Research Center CRC 1044, FOR5327, GRK 2149; Istituto Nazionale di Fisica
Nucleare, Italy; Ministry of Development of Turkey under Contract No.
DPT2006K-120470; National Research Foundation of Korea under Contract No.
NRF-2022R1A2C1092335; National Science and Technology fund of Mongolia;
National Science Research and Innovation Fund (NSRF) via the Program
Management Unit for Human Resources \& Institutional Development, Research and
Innovation of Thailand under Contract No. B16F640076; Polish National Science
Centre under Contract No. 2019/35/O/ST2/02907; The Swedish Research Council;
U. S. Department of Energy under Contract No. DE-FG02-05ER41374